\documentclass[showpacs]{revtex4}
\usepackage{amsmath}
\usepackage{amssymb}
\usepackage{graphicx}

\begin{document}

\title{Phantom thick brane in 5D bulk}
\author{Vladimir Dzhunushaliev
\footnote{Senior Associate of the Abdus Salam ICTP}}
\email{dzhun@krsu.edu.kg}
\affiliation{Dept. Phys. and Microel.
Engineer., Kyrgyz-Russian Slavic University, Bishkek, Kievskaya Str.
44, 720021, Kyrgyz Republic}

\author{Vladimir Folomeev}
\email{vfolomeev@mail.ru}
\affiliation{Institute of Physics of NAS
KR, 265 a, Chui str., Bishkek, 720071,  Kyrgyz Republic}

\author{Shynaray Myrzakul and Ratbay Myrzakulov}
\email{cnlpmyra1954@yahoo.com, cnlpmyra@mail.ru}
\affiliation{Dept. Gen. and Theor. Phys., Eurasian National University, Astana, 010008, Kazakhstan}

\begin{abstract}
A model of a thick brane in 5D bulk supported by two phantom scalar fields is considered. The comparison with a thick brane supported by two usual scalar fields is carried out. The distinctions between a thick brane supported by one usual scalar field and our model have been pointed out.
\end{abstract}


\maketitle

\section{Introduction}
A discovery of the accelerated expansion of the Universe \cite{Riess_a,Riess_b,Star_de_a,Star_de_b,Star_de_c} and a
great deal of work attempting to  model this phenomenon allows
to suppose that existence of a special class of scalar fields -- phantom fields -- is possible in the Universe~\cite{Star_172}. The phantom fields are
some type of matter with the violated weak or null energy conditions.
One of the ways of introduction of such fields is consideration of theories of a scalar
field with a negative sign before a kinetic term (ghost fields)~\cite{Szydlowski:2004jv,Elizalde,Capozziello}. 
In~\cite{Dzhunushaliev:2007cs}, using the ghost fields,
we have obtained regular wormhole, particle-like and brane-like solutions. Another way consists in a consideration of braneworld models of dark energy
which also allow violation of the weak or null energy conditions~\cite{Sahni_bw_de_a,Sahni_bw_de_b}.

The accelerated expansion of the Universe is being supported in phantom models by the phantom fields filling all the Universe.
In this paper we suppose that the phantom field really fills the Universe, and we use this assumption for obtaining
a thick-brane solution in a 5D spacetime with two interacting phantom scalar fields. Previously, one of us has already obtained a similar
regular solution for two usual interacting  scalar fields~\cite{Dzhunushaliev:2006vv}. Here we want to compare the solutions
from~\cite{Dzhunushaliev:2006vv} with a solution to be found below, and to clarify the difference  between
thick-brane solutions with phantom and usual scalar fields.

Also we want to compare our results with the results from  \cite{Bronnikov:2003gg} where
 thick branes with one usual scalar field are under consideration (see also earlier work
 on thick branes~\cite{Kobayashi:2001jd}). It was shown there that
corresponding solutions have AdS asymptotics, the potential
 $V(\varphi)$ must be an alternating function, and also some fine tuning condition should be satisfied.

\section{Equations and solutions}

We start from the Lagrangian
\begin{equation}
\label{Lagr_phantom}
	L=-\frac{R}{2}+\epsilon\left[\frac{1}{2}\partial_A\varphi\partial^A\varphi+
	\frac{1}{2}\partial_A\chi\partial^A\chi-V(\varphi,\chi)\right],
\end{equation}
where capital Latin indices run over $0,1,2,3,5$, $\epsilon=+1$ for the usual scalar fields, and
 $\epsilon=-1$ for the phantom ones;  $\varphi,\chi$ are two interacting  scalar fields with the potential
\begin{equation}
\label{pot_phantom}
    V(\varphi,\chi)=\frac{\Lambda_1}{4}(\varphi^2-m_1^2)^2+
    \frac{\Lambda_2}{4}(\chi^2-m_2^2)^2+\varphi^2 \chi^2-V_0,
\end{equation}
where $V_0$ is some constant. The corresponding energy-momentum tensor is
\begin{equation}
\label{emt_phantom}
	T^B_A=\epsilon \left\{
	\partial_A\varphi\partial^A\varphi+
	\partial_A\chi\partial^A\chi-\delta^B_A\left[\frac{1}{2}\partial_A\varphi\partial^A\varphi+
	\frac{1}{2}\partial_A\chi\partial^A\chi-V(\varphi,\chi)\right]
	\right\}.
\end{equation}
We consider a problem when all variables depend only on the extra coordinate $r$. Then metric for a 4D-brane embedded
in an external five-dimensional spacetime can be written in the form
\begin{equation}
\label{metr_phantom}
	ds^2=a(r)^2\left(dt^2-dx^2-dy^2-dz^2\right)-dr^2.
\end{equation}
Using this metric, the Einstein equations will be
\begin{eqnarray}
\label{einst_phantom1}
\frac{a^{\prime\prime}}{a}-\left(\frac{a^\prime}{a}\right)^2&=&-\frac{\epsilon}{3}\left(\varphi^{\prime 2}+\chi^{\prime 2}\right),\\
-6\left(\frac{a^\prime}{a}\right)^2&=&\epsilon\left[-\frac{1}{2}\left(\varphi^{\prime 2}+\chi^{\prime 2}\right)+V(\varphi,\chi)\right],
\label{einst_phantom2}
\end{eqnarray}
where the first equation was obtained by summation of
$\left(^t_t\right)$ and $\left(^r_r\right)$ components.

The equations for the scalar fields are
\begin{eqnarray}
\label{field_phantom1}
\varphi^{\prime\prime}+4\frac{a^\prime}{a}\varphi^{\prime}&=&\varphi\left[2\chi^2+\Lambda_1\left(\varphi^2-m_1^2\right)\right],\\
\chi^{\prime\prime}+4\frac{a^\prime}{a}\chi^{\prime}&=&\chi\left[2\varphi^2+\Lambda_2\left(\chi^2-m_2^2\right)\right].
\label{field_phantom2}
\end{eqnarray}
Let us examine the system of equations
 \eqref{einst_phantom1}-\eqref{field_phantom2} with
 boundary conditions
\begin{eqnarray}
	a(0) &=& a_0 ,
\label{bound_a}\\
	a'(0) &=& 0 ,
\label{bound_a_prime}\\
	\varphi(0) &=& \varphi_0 , \quad \varphi^{\prime}(0) = 0 ,
\label{bound_phi}\\
	\chi(0) &=& \chi_0 , \quad \chi'(0) = 0.
\label{bound_chi}
\end{eqnarray}
The conditions
 \eqref{bound_a_prime}-\eqref{bound_chi} and equation \eqref{einst_phantom2} give the following value of the constant
$V_0$:
\begin{equation}
\label{V0_phantom}
V_0 = \frac{\Lambda_1}{4} \left(\varphi^2_0 - m_1^2\right)^2 +
\frac{\Lambda_2}{4} \left(\chi^2_0 - m_2^2\right)^2 +  \varphi^2_0 \chi^2_0.
\end{equation}

Let us choose the following numeric parameters for numerical analysis:
\begin{equation}
	a_0 = \varphi_0 = 1, \quad
	\chi_0 = \sqrt{0.6}, \quad
	\Lambda_1 = 0.1, \quad
	\Lambda_2 = 1.0 .
\end{equation}

\begin{figure}[h]
\begin{center}
 \includegraphics[width=12cm]{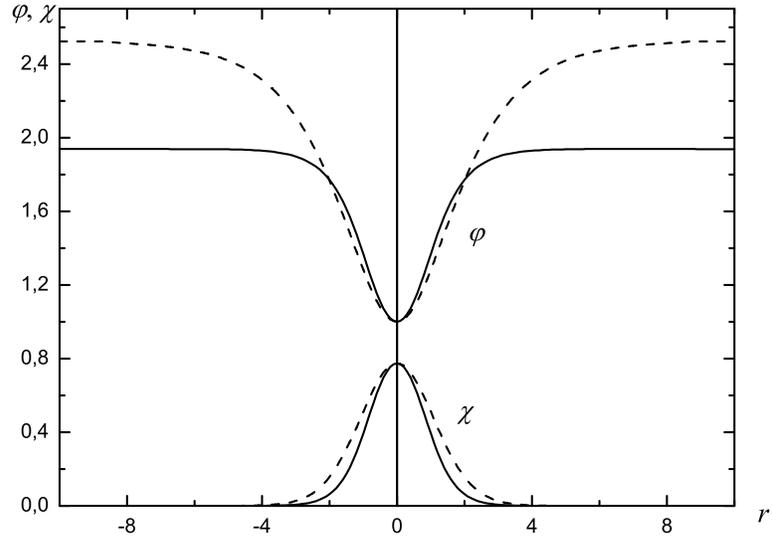}
  \vspace{-1cm}
\caption{The scalar fields profiles for the cases
$\epsilon=-1$ (solid lines) and $\epsilon=+1$ (dashed lines).}
 \label{fig:fields_phantom}
 \end{center}
\end{figure}

\begin{figure}[h]
\begin{center}
 \includegraphics[width=12cm]{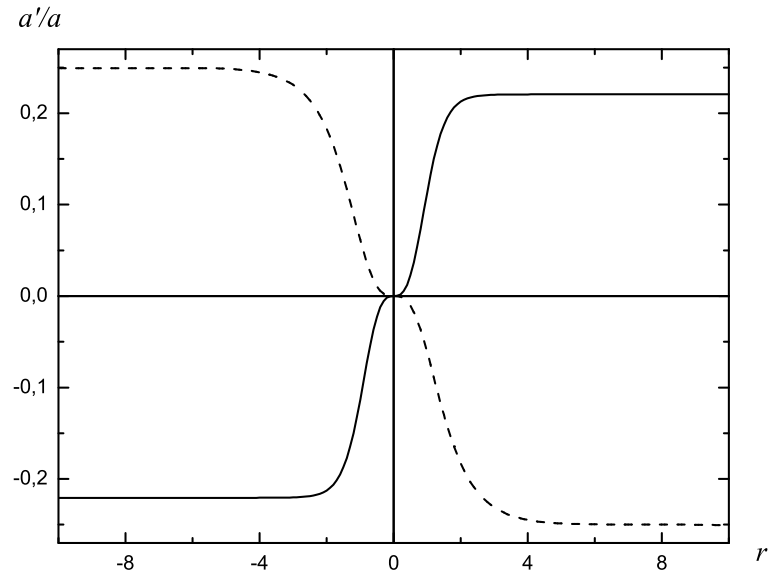}
  \vspace{-1cm}
\caption{The profiles of the functions $\mathcal{H}=a^\prime/a$ for the cases
$\epsilon=-1$ (solid line) and $\epsilon=+1$ (dashed line).}
 \label{fig:metric_phantom}
 \end{center}
\end{figure}

\begin{figure}[h]
\begin{center}
 \includegraphics[width=12cm]{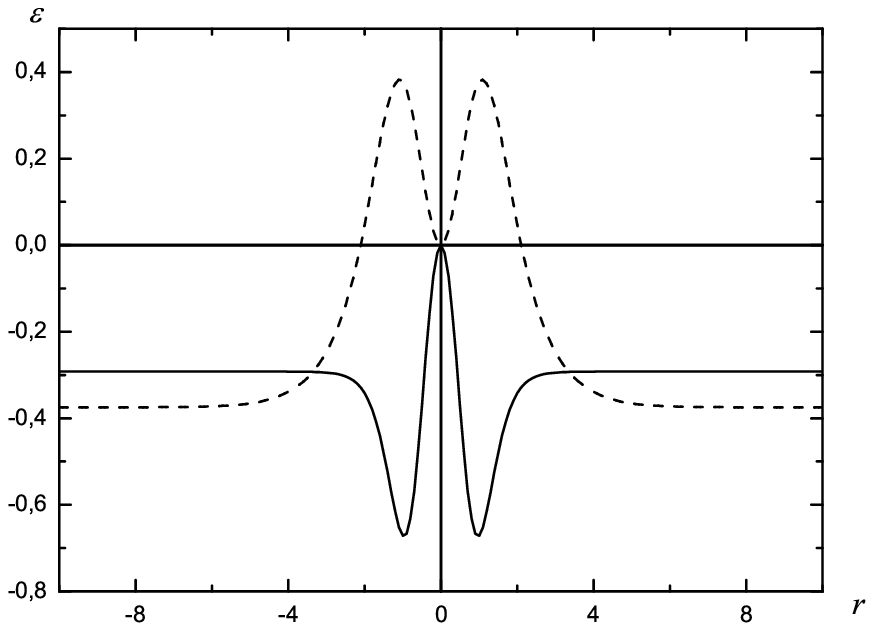}
  \vspace{-1cm}
\caption{The energy density $\varepsilon$ for the cases
$\epsilon=-1$ (solid line) and $\epsilon=+1$ (dashed line).}
 \label{fig:energy_phantom}
 \end{center}
\end{figure}

The model with $\epsilon=+1$ was considered in
\cite{Dzhunushaliev:2006vv}.  Here we examine a case when
$\epsilon=-1$, and compare obtained results with the results from
\cite{Dzhunushaliev:2006vv}. At numerical analysis of the system
\eqref{einst_phantom1}-\eqref{field_phantom2} one should solve a nonlinear problem for calculation of
eigenvalues of the parameters  $m_1$ and $m_2$. This problem is being solved by the shooting method which is described
in detail in \cite{Dzhunushaliev:2006vv}. The obtained results are presented in Figs.~\ref{fig:fields_phantom}-\ref{fig:energy_phantom}.
The solutions for the case $\epsilon=-1$ are shown by solid lines, and for the case  $\epsilon=+1$ -- by dashed lines.
The following values of the masses were found: $m_1\approx 1.93927$ and
$m_2\approx 1.97696852$ for the case $\epsilon=-1$, $m_1\approx 2.5220566669937$ and
$m_2\approx 1.85622511328021$ for $\epsilon=+1$.

Let us estimate asymptotic behavior of the solutions. One can see from
 \eqref{einst_phantom2} that asymptotically
\begin{eqnarray}
a_{(-1)}&\approx& a_\infty e^{k_{(-1)} r}, \quad k_{(-1)}=\sqrt{\frac{1}{6}\left(\frac{\Lambda_2}{4}m_2^4-V_0\right)}>0,\\
a_{(+1)}&\approx& a_\infty e^{-k_{(+1)} r}, \quad k_{(+1)}=\sqrt{-\frac{1}{6}\left(\frac{\Lambda_2}{4}m_2^4-V_0\right)}>0,
\end{eqnarray}
where $a_\infty$ is an asymptotic value of the metric function, and subscripts
 (-1) and (+1) refer to cases with phantom and usual fields, respectively. Then, using these expressions and
 looking for a solution of equations
\eqref{field_phantom1}-\eqref{field_phantom2} in the form
$$
\varphi \approx m_1-\delta \varphi, \quad \chi \approx \delta \chi,
$$
we have the following asymptotic form for the scalar fields:
\begin{eqnarray}
	\delta \varphi_{(-1)}&=&\varphi_\infty 	\exp{\left[-\left(2k_{(-1)}+
	\sqrt{4k_{(-1)}^2+ 2	\Lambda_1m_1^2}\right)r\right]},
\\
	\delta \chi_{(-1)}&=&\chi_\infty \exp{\left[\left(2k_{(-1)}-
	\sqrt{4k_{(-1)}^2+2m_1^2-\Lambda_2 m_2^2}\right)r\right]},
\\
	\delta \varphi_{(+1)}&=&\varphi_\infty \exp{\left[\left(2k_{(+1)}-
	\sqrt{4k_{(+1)}^2+ 2 \Lambda_1m_1^2}\right)r\right]},
\\
	\delta \chi_{(+1)}&=&\chi_\infty \exp{\left[-\left(2k_{(+1)}+
	\sqrt{4k_{(+1)}^2+2m_1^2-\Lambda_2 m_2^2}\right)r\right]},
\end{eqnarray}
where $\varphi_\infty, \chi_\infty$ are asymptotical values of the functions. One can see from the obtained expressions that
asymptotically
 $\delta \varphi_{(\pm1)} \rightarrow 0$ and
$\delta \chi_{(\pm1)} \rightarrow 0$, i.e. the solutions are finite.

In Fig.~\ref{fig:energy_phantom} the profiles of the energy density
\begin{equation}
	\varepsilon(r)=\epsilon\left[\frac{1}{2}\varphi^{\prime 2}+\frac{1}{2}\chi^{\prime 2}+V(\varphi,\chi)\right]
\label{energy-density}
\end{equation}
for the phantom and usual scalar fields are shown.

\section{Stability analysis}

Let us now study the dynamical stability of the above solutions against linear perturbations.
To do this we have to write field equations following from the Lagrangian \eqref{Lagr_phantom}
keeping time dependence of metric and scalar fields functions. Then $\left(^t_t\right), \left(^r_r\right)$ and
$\left(^r_t\right)$ components of the Einstein equations will be:
\begin{eqnarray}
-3\left[-\frac{\dot{a}^2}{a^4}+\left(\frac{a^\prime}{a}\right)^2+\frac{a^{\prime\prime}}{a}\right]&=&
\epsilon\left[\frac{1}{2}a^2\left(\dot{\varphi}^2+\dot{\chi}^2\right)+\frac{1}{2}\left(\varphi^{\prime 2}+\chi^{\prime 2}\right)+V\right],\\
-3\left[-\frac{\ddot{a}}{a^3}+2\left(\frac{a^\prime}{a}\right)^2\right]&=&
\epsilon\left[-\frac{1}{2}a^2\left(\dot{\varphi}^2+\dot{\chi}^2\right)-\frac{1}{2}\left(\varphi^{\prime 2}+\chi^{\prime 2}\right)+V\right],\\
3\left[-\frac{\dot{a}a^\prime}{a^2}+\frac{\dot{a}^\prime}{a}\right]&=&
\epsilon\left(-\dot{\varphi}\varphi^\prime-\dot{\chi}\chi^\prime\right),
\end{eqnarray}
where differentiation with respect to $t$ and $r$ is denoted by
a $dot$ and a $prime$ correspondingly.

Introducing a new metric function $\alpha$ via
$$a=e^\alpha,$$
%one can derive for difference
%$G_t^t-G_r^r$:
%\begin{equation}
%\label{diff}
%\alpha^{\prime\prime}+e^{-2\alpha}\ddot{\alpha}=-\frac{\epsilon}{3}\left[e^{2\alpha}\left(\dot{\varphi}^2+\dot{\chi}^2\right)+
%\left(\varphi^{\prime 2}+\chi^{\prime 2}\right)\right].
%\end{equation}
the component $\left(^r_t\right)$ takes the form
\begin{equation}
\label{mixed}
3\dot{\alpha}^\prime=\epsilon\left(-\dot{\varphi}\varphi^\prime-\dot{\chi}\chi^\prime\right).
\end{equation}
The corresponding equations for the scalar fields are
\begin{eqnarray}
\label{sfe_phi}
e^{-2\alpha}\left[\ddot{\varphi}+2\dot{\alpha}\dot{\varphi}\right]-\varphi^{\prime\prime}-4\alpha^{\prime}\varphi^{\prime}=
-\varphi\left[2\chi^2+\Lambda_1\left(\varphi^2-m_1^2\right)\right],\\
\label{sfe_chi}
e^{-2\alpha}\left[\ddot{\chi}+2\dot{\alpha}\dot{\varphi}\right]-\chi^{\prime\prime}-4\alpha^{\prime}\chi^{\prime}=
-\chi\left[2\varphi^2+\Lambda_2\left(\chi^2-m_2^2\right)\right].
\end{eqnarray}

We perturb the solutions of the system \eqref{mixed}-\eqref{sfe_chi} by expanding the metric function and scalar
fields functions to first order as follows
\begin{eqnarray}
\alpha&=&\alpha_0(r)+\alpha_1(r)\cos{\omega t},\\
\varphi&=&\varphi_0(r)+\varphi_1(r)\cos{\omega t},\\
\chi&=&\chi_0(r)+\chi_1(r)\cos{\omega t},
\end{eqnarray}
where the index $0$ indicates the static background solutions of equations \eqref{einst_phantom1}-\eqref{field_phantom2},
and the index $1$ refers to perturbations. Then one has
from \eqref{mixed}:
\begin{equation}
\label{mixed_pert}
\alpha_1^\prime=-\frac{\epsilon}{3}\left(\varphi_0^\prime
\varphi_1+\chi_0^\prime\chi_1\right).
\end{equation}
%1) from \eqref{diff}:
%\begin{equation}
%\label{diff_pert}
%\alpha_1^{\prime\prime}-\omega^2 e^{-2\alpha_0}\alpha_1=-\frac{2\epsilon}{3}\left(\varphi_0^\prime
%\varphi_1^\prime+\chi_0^\prime\chi_1^\prime\right);
%\end{equation}
Using the last expression, one can rewrite equations \eqref{sfe_phi}, \eqref{sfe_chi} as
\begin{eqnarray}
\label{sfe_phi_pert}
\varphi_1^{\prime\prime}+4\alpha_0^\prime\varphi_1^\prime-
\left[\frac{4\epsilon}{3}\varphi_0^{\prime 2}+\Lambda_1\left(2\varphi_0^2-m_1^2\right)\right]\varphi_1-
\left(\frac{4\epsilon}{3}\varphi_0^\prime\chi_0^\prime+4\varphi_0\chi_0\right)\chi_1+\omega^2 e^{-2\alpha_0}\varphi_1&=&0,\\
\chi_1^{\prime\prime}+4\alpha_0^\prime\chi_1^\prime-
\left[\frac{4\epsilon}{3}\chi_0^{\prime 2}+\Lambda_2\left(2\chi_0^2-m_2^2\right)\right]\chi_1-
\left(\frac{4\epsilon}{3}\varphi_0^\prime\chi_0^\prime+4\varphi_0\chi_0\right)\varphi_1+\omega^2 e^{-2\alpha_0}\chi_1&=&0.
\label{sfe_chi_pert}
\end{eqnarray}

\begin{figure}[h]
\begin{center}
 \includegraphics[width=12cm]{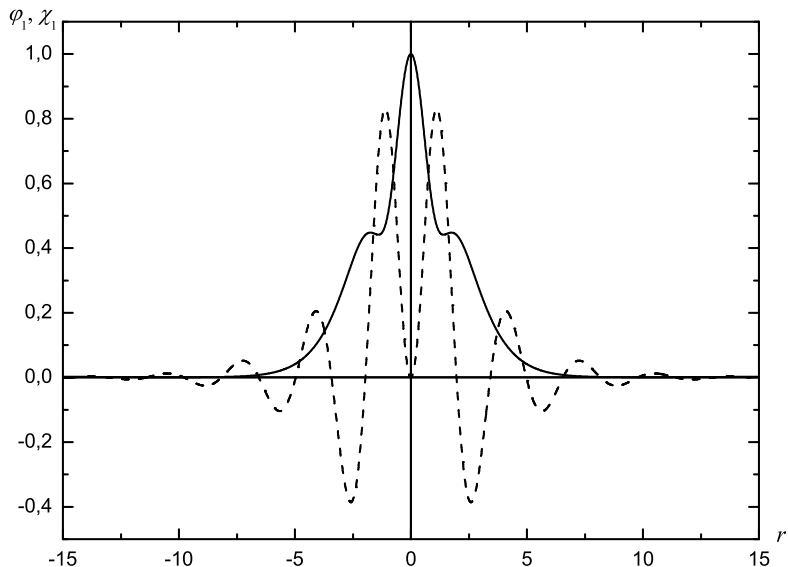}
  \vspace{-1cm}
	\caption{The phantom scalar fields perturbations $\varphi_1$
(solid line) and $\chi_1$ (dashed line) from equations \eqref{sfe_phi_pert}, \eqref{sfe_chi_pert} with the
boundary conditions \eqref{bound_pert} and $\omega^2\approx 1.6419$.}
 \label{fig:perturb_phant}
 \end{center}
\end{figure}

For existence of stable solutions it is necessary to provide positiveness of an eigenvalue $\omega^2$.
If there is a negative eigenvalue $\omega^2$ then the
solution will be unstable since then $\varphi_1,\chi_1 \sim e^{i \omega t}$ will grow exponentially.
In order to make clear this question, we will search for numerical solutions of equations \eqref{sfe_phi_pert}, \eqref{sfe_chi_pert}
with
boundary conditions
\begin{equation}
\label{bound_pert}
\varphi_1(0)=1, \quad \varphi_1^\prime(0)=0, \quad \chi_1(0)=0, \quad \chi_1^\prime(0)=0.
\end{equation}
For a case $\epsilon=-1$ one can find that at $\omega^2\approx 1.6419$ the following regular solutions for $\varphi_1, \chi_1$ exist
(see Fig. \ref{fig:perturb_phant}).
It indicates that the solutions with phantom scalar fields are stable for the case under consideration.
But for a case $\epsilon=1$ we could not find any regular solutions of equations \eqref{sfe_phi_pert}, \eqref{sfe_chi_pert}.
It seems that there is no stable solutions for usual scalar fields (at least for the model's parameters we used here).

\section{Discussion and conclusions}

In our opinion, the main distinction of the phantom thick brane from a brane created by usual scalar fields consists in asymptotical
behavior of the metric function $a(r)$: for the phantom thick brane $a(r)~\approx~a_\infty e^{k_{(-1)} r} \rightarrow \infty$, and
for the usual brane $a(r)~\approx~a_\infty e^{-k_{(+1)} r} \rightarrow 0$. We think that it happens due to the following reasons:
let a solution has the following form near $r=0$:
\begin{eqnarray}
	a(r) &=& a_0 + a_4 \frac{r^4}{12} + \cdots ,
\label{3-10}\\
	\varphi(r) &=& \varphi_0 + \varphi_2 \frac{r^2}{2} + \cdots ,
\label{3-20}\\
	\chi(r) &=& \chi_0 + \chi_2 \frac{r^2}{2} + \cdots .
\label{3-30}
\end{eqnarray}
Their substitution in \eqref{einst_phantom1} gives the following result
\begin{equation}
	a_4 = - \frac{\epsilon}{3} \left( \varphi_2^2 + \chi_2^2 \right).
\label{3-40}
\end{equation}
It implies that at $\epsilon < 0$, $a_4 > 0$ and one has (at least near $r=0$) a steadily increasing function which was confirmed by
the numerical calculations presented in Fig.~\ref{fig:metric_phantom}. At  $\epsilon > 0$, $a_4 < 0$ and one has a decreasing function.

Another interesting feature of the obtained solution is that the energy density of the phantom brane at the origin  of coordinates $r=0$
has an absolute maximum. At the same time, the usual brane has a local minimum at $r=0$. Physically, it implies that in the phantom case the
brane looks like a hill on the background of phantom matter with constant energy density. In case of usual brane a maximum of energy
is concentrated near the brane on both sides. In our opinion, it may results in different scenarios of trapping of 4D matter on the phantom and
usual branes.

\begin{figure}[h]
\begin{center}
 \includegraphics[width=12cm]{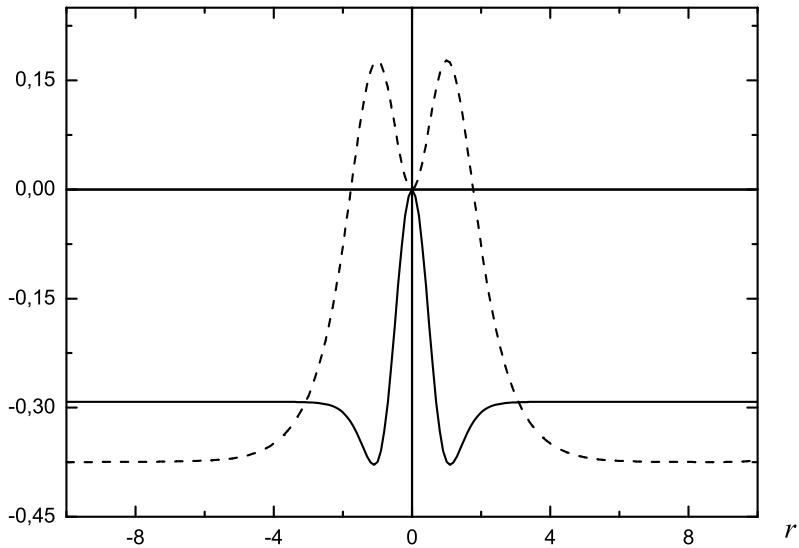}
  \vspace{-1cm}
	\caption{The profile of the potential energy $\epsilon V(\varphi, \chi)$ for the cases
$\epsilon=-1$ (solid line) and $\epsilon=+1$ (dashed line).}
 \label{fig:potential}
 \end{center}
\end{figure}

At comparison of our  solutions with the results from \cite{Bronnikov:2003gg} one might note the following differences:
(a) the metric function $a(r)$ increases exponentially;
(b) the potential $V(r)=V(\varphi(r), \chi(r))$ does not change a sign
(see Fig.~\ref{fig:potential});
(c)~in consequence of the condition (b), there is no a fine tuning condition.

In Ref.~\cite{Bronnikov:2003gg} such fine tuning condition is necessary for obtaining of a thick brane supported by one scalar field.
This fine tuning condition consists in the following: using the metric
\begin{equation}	
	ds_5^2 = e^{2F(z)} \eta_{\mu \nu} dx^\mu dx^\nu - e^{8F(z)} dz^2
\label{3-45}
\end{equation}
with the harmonic coordinate $z$, such that $\sqrt{g} g^{zz}=-1$,  one should has the following fine tuning condition
\begin{equation}	
	\bar{V}(\infty) = 0,
\label{3-47}
\end{equation}
where
\begin{equation}	
	\bar{V}(z) = \int \limits_{0}^{z} \sqrt{g} V(z_1)\, dz_1 =
   \int_{0}^{z}\! e^{8F(z_1)} V(z_1)\,dz_1.
\label{3-49}
\end{equation}
In our case the corresponding condition could be obtained as follows: from equations \eqref{einst_phantom1}, \eqref{einst_phantom2} one has
\begin{equation}	
	a^\prime (r) = -\epsilon \int \limits_0^r a(r)
	V \Bigl( \varphi(r), \chi(r) \Bigl) dr,
\label{3-50}
\end{equation}
and because of $a^\prime (\infty) \neq 0$, we have
\begin{equation}	
	\int \limits_0^\infty a(r) 	V \Bigl( \varphi(r), \chi(r) \Bigl) dr \neq 0.
\label{3-60}
\end{equation}
This indicates that the nontrivial potential $V(\varphi, \chi)$ does not change a sign as it can be seen from Fig.~\ref{fig:potential}.

So we have obtained the solutions describing a thick phantom brane in 5D bulk. There are two main distinctive features of these solutions:
a) they have different asymptotic behavior; b) the solutions with phantom scalar fields are stable, and the solutions with usual scalar
fields are not (at least for the parameters $\Lambda_1, \Lambda_2, \varphi(0), \chi(0)$ we used here).

\end{document}